\documentclass[10pt,pra,aps,twocolumn,superscriptaddress,floatfix,showpacs]{revtex4-2}
\usepackage[utf8]{inputenc}
\usepackage[american,british]{babel}
\usepackage[T1]{fontenc}
\usepackage[pdftex]{graphicx}  
\usepackage{graphicx, xcolor}
\usepackage{dcolumn}
\usepackage{braket}
\usepackage{bm}
\usepackage{amsmath,amsthm,amssymb}
\usepackage{color}
\usepackage{verbatim}
\usepackage[noend]{algpseudocode}
\usepackage[plain,ruled,vlined]{algorithm2e}
\usepackage{ulem}
\usepackage{appendix} 
\usepackage{hyperref}
\hypersetup{
 colorlinks=true,
 linkcolor=blue,
 anchorcolor = blue,
 citecolor = blue,
 filecolor = blue,
 urlcolor = blue
}

\definecolor{green}{RGB}{0,110,0}
\definecolor{blue}{RGB}{0,0,130}

\newcommand{\vv}[1]{#1}

\usepackage{stmaryrd}

\usepackage{tikz}
\usetikzlibrary{patterns}
\usetikzlibrary{positioning, arrows.meta}
\usetikzlibrary{shapes.misc}
\usetikzlibrary{fit}
\usetikzlibrary{calc,overlay-beamer-styles}

\algnewcommand{\IfThenElse}[3]{%
  \State \algorithmicif\ #1\ \algorithmicthen\ #2\ \algorithmicelse\ #3}
\algnewcommand{\IfThen}[2]{%
  \State \algorithmicif\ #1\ \algorithmicthen\ #2}

\usepackage{lipsum}
\usepackage{soul}
\makeatletter

\newcommand{\Graph}{{\mathcal{G}}}

\newcommand{\vertices}{{\mathcal{V}}}

\definecolor{darkgreen}{RGB}{15, 30, 35}
\definecolor{softgreen}{RGB}{23, 48, 53}
\definecolor{brightgreen}{RGB}{225, 246, 233}
\definecolor{mintgreen}{RGB}{0, 200, 135}
\definecolor{metalblue}{RGB}{57, 115, 120}
\definecolor{neonblue}{RGB}{146, 200, 229}
\definecolor{pasqalpurple}{RGB}{134, 123, 250}
\definecolor{pasqalorange}{RGB}{255, 152, 110}

\begin{document}
\title{Quantum Optimization on Rydberg Atom Arrays with Arbitrary Connectivity: Gadgets Limitations and a Heuristic Approach} 
\date{\today}

\author{Pierre Cazals}
\affiliation{Pasqal, 24 rue Emile Baudot, 91120 Palaiseau, France}
\author{Amalia Sorondo}
\affiliation{Pasqal, 24 rue Emile Baudot, 91120 Palaiseau, France}
\author{Victor Onofre}

\affiliation{Pasqal, 24 rue Emile Baudot, 91120 Palaiseau, France}
\author{Constantin Dalyac}
\affiliation{Pasqal, 24 rue Emile Baudot, 91120 Palaiseau, France}
\author{Wesley Coelho}
\affiliation{Pasqal, 24 rue Emile Baudot, 91120 Palaiseau, France}
\author{Vittorio Vitale}
\email{vittorio.vitale@pasqal.com}
\affiliation{Pasqal, 24 rue Emile Baudot, 91120 Palaiseau, France}

\begin{abstract}
Programmable quantum systems based on Rydberg atom arrays have recently emerged as a promising testbed for combinatorial optimization. Indeed, the Maximum Weighted Independent Set problem on unit-disk graphs can be efficiently mapped to such systems due to their geometric constraints. However, extending this capability to arbitrary graph instances typically necessitates the use of reduction gadgets, which introduce additional experimental overhead and complexity. Here, we analyze the complexity‐theoretic limits of polynomial reductions from arbitrary graphs to unit‐disk instances. We prove any such reduction incurs a quadratic blow‐up in vertex count and degrades solution approximation guarantees. As a practical alternative, we propose a divide‐and‐conquer heuristic with only linear overhead which leverages precalibrated atomic layouts. We benchmark it on Erd\"os-Rényi graphs, and demonstrate feasibility on the Orion Alpha processor.
\end{abstract}

\maketitle
\begin{figure*} 
\includegraphics[width=1.\linewidth]{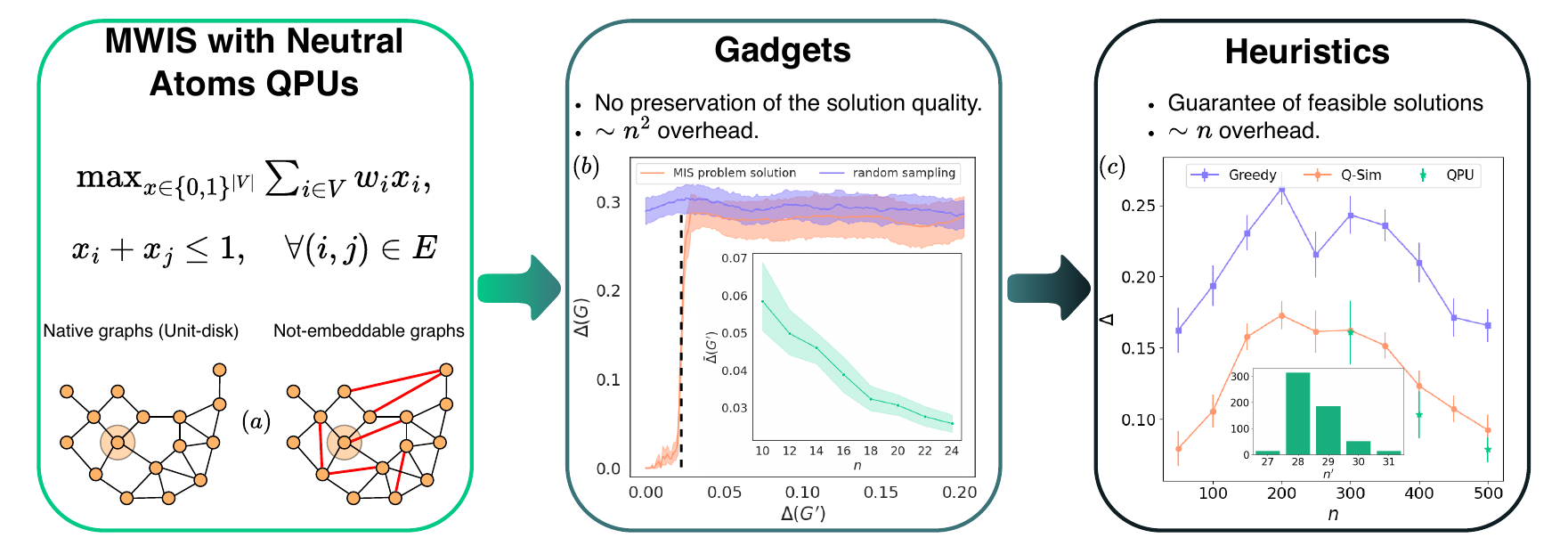}
    \caption{{\it Summary of the paper}. We solve the Maximum Weighted Independent Set (MWIS) problem on Rydberg Atoms arrays. $(a)$ The native graphs that can be embedded in the machine are Unit-Disk (UD) graphs, for which a Polynomial-Time Approximation Scheme (PTAS) exists. MWIS problems on UD graphs exemplify an `easy subset' of NP-hard problems to solve for classical algorithms. 
    Generic not embeddable graphs exhibit long-range couplings that violate the UD constraints. In the example, the edges that do not comply with the UD constraint are highlighted in red. To tackle the MWIS on arbitrary graphs one must employ reductions that allow to embed it on native layouts. Other than incurring in a quadratic overhead in mapping from the original graph $G$ to $G'$, such reductions do not preserve the solution quality. $(b)$ Solutions in $G'$ with gaps $\Delta(G')$ larger than a threshold $\bar{\Delta}(G')$ (dashed vertical line) map to poor quality solutions of the problem in $G$ -- comparable to random sampling. $\bar{\Delta}(G')$ goes to zero $\propto 1/n$. $(c)$ We propose a heuristic algorithm for solving the MWIS problem. \vv{This approach splits arbitrary graphs into smaller instances of the MWIS on precalibrated atomic layouts, it guarantees feasible solutions and incurs an $\mathcal{O}(n)$ overhead}. We plot the gap to the best solution computed by means of CPLEX~\cite{cplex2024} and Simulated annealing (SA)~\cite{dimod} We show results of both quantum simulations (Q-sim, in orange) and experiments (QPU, in green) by employing our method, in comparison with the solution obtained by a Greedy algorithm (purple). We consider Erd\"os-Rényi graphs of size up to $n=500$ and $p=0.5$ (probability of drawing an edge between two nodes). The simulation results are averaged over 100 instances while the experimental results over 10. The error bars correspond to the standard deviation of the mean. In the inset of $(c)$, we show an histogram counting the occurrences of the largest subgraph obtained with our method as a function of its size $n'$.}
    \label{fig:fig1}
\end{figure*}

Combinatorial optimization problems, which require selecting optimal solutions from a finite --- often exponentially large --- set of possibilities, appear throughout logistics, telecommunications, finance, and healthcare~\cite{paschos2014applications,resende2003combinatorial,du2019real,ali2015mathematical,leclerc2025}. Among them, a canonical NP-hard~\cite{van1991handbook} example is the Maximum Weighted Independent Set (MWIS) problem: given a graph $G = (V, E)$ with vertex weights $w_i$, the task is to find a subset $S \subseteq V$ such that no two vertices in $S$ are adjacent, and the total weight $\sum_{i \in S} w_i$ is maximized~\footnote{When $w_i=1$, $\forall\; i$, one has the unweighted version of the problem called simply Maximum Independent Set (MIS) problem.}.

Neutral atom quantum processing units (QPUs) have recently enabled the physical realization of the MWIS on unit-disk (UD) graphs, where vertices correspond to atoms in two-dimensional space and edges represent proximity within a fixed interaction radius. The Rydberg blockade mechanism enforces the independence constraint by suppressing simultaneous excitations of nearby atoms~\cite{lukin2001dipole,browaeys2020many,saffman2010quantum,urban2009observation,labuhn2016tunable,ebadi2021quantum,dalyac2021qualifying,lucas2014ising,pichler2018quantum,byun2022quantum}. This native encoding has made UD-MWIS a key benchmark for quantum optimization experiments (Fig.~\ref{fig:fig1}a).

However, most real-world problems exhibit non-local connectivity that cannot be directly embedded into a UD geometry. To bridge this gap, various polynomial-time reduction techniques -- often called \textit{gadgets} -- have been proposed to encode general graphs into UD form~\cite{kim2022rydberg,byun2022quantum,dalyac2023exploring,nguyen2023quantum,lanthaler2023rydberg,byun2024rydberg,jeong2023quantum}. These reductions create a one to one mapping of the optimal solution of the original problem to the UD-reduced one, but incur in substantial overhead in qubit count and calibration complexity.

Yet, in practical settings, obtaining the exact ground state --- which encodes the optimal solution --- of such embedded problems is often unfeasible: even if the reduction is exact, solving adiabatically the resulting instance is known to require exponential time in the system size due to spectral gap closing~\cite{bombieri2025quantum}. As a result, approximate solutions become the realistic goal—especially for large-scale instances.

In this work, we identify two fundamental theoretical limitations of such deterministic reduction schemes:
$(i)$ under the Exponential Time Hypothesis (ETH)~\cite{impagliazzo2001problems}, any polynomial-time mapping from general graphs to UD-graphs must introduce a quadratic blow-up in the number of vertices, $\mathcal{O}(n^{2})$;
$(ii)$ More critically, such reductions do not guarantee preservation of approximate solutions: a small optimality gap in the UD-reduced instance does not imply any meaningful performance on the original problem. Our numerical simulations confirm this degradation: near-optimal solutions in the embedded problem often map back to solutions with quality close to that of random sampling (Fig.~\ref{fig:fig1}b).

These results are hardware-agnostic and apply universally to any deterministic embedding strategy, highlighting intrinsic limitations of general-purpose gadget-based mappings. The root cause lies in the structure of the reductions: embedding non-local edges often requires inserting long induced paths of auxiliary vertices. These chains inflate the instance size and introduce numerous local minima in the solution landscape, many of which correspond to unfeasible or poor-quality solutions in the original problem.

Given these limitations, the path forward lies in heuristic methods that are tailored to problem structure. Rather than relying on general-purpose reductions, such approaches exploit minor embedding to guide the encoding process or restrict the search space. For instance, structured gadget constructions~\cite{boothby2016fast, ender2023parity, ParityQC2024} aim to preserve useful connectivity patterns and reduce degeneracy in the solution landscape. In this paper, we introduce a recursive heuristic that avoids full-graph embedding. Instead, our method iteratively extracts embeddable subgraphs, solves them independently, and merges their solutions into a valid independent set for the original problem. By operating only on subgraphs that are natively compatible with the current generation of the hardware, and relying on pre-calibrated atomic configurations, our method supports near-term experimental realization.

Appendices provide implementation details of the proposed heuristic, discussion of the quantum algorithm used on Pasqal's Orion Alpha QPU, and additional numerical results.

\section{Reductions drawbacks}
\label{sec:reduction_drawbacks}

\subsection{Quadratic overhead}
First, let us turn to complexity theory to give theoretical bounds to the application of polynomial reductions for optimization purposes. In App.~\ref{sec:asymptotic_notation} we summarize the asymptotic notation for the unfamiliar reader.

A standpoint for deriving tight conditional lower bounds for approximation algorithms and exact exponential‐time procedures is the Exponential Time Hypothesis (ETH)~\cite{impagliazzo2001problems}.
The latter conjectures that no deterministic algorithm can solve 3‐SAT~\footnote{The 3-SAT (three-satisfiability) problem is the decision problem of determining whether a given Boolean formula in conjunctive normal form -- where each clause contains exactly three literals -- admits a truth assignment that satisfies all clauses. It is the canonical NP-complete problem, meaning that it is both in NP and that every problem in NP can be reduced to it in polynomial time.} in time significantly better than $2^{\Theta(n)}$, where $n$ is the number of variables.
Under ETH, suppose we have two decision problems $\mathcal{A}$ and $\mathcal{B}$:
\begin{itemize}
  \item $\mathcal{A}$ admits a sub-exponential time algorithm running in time \vv{$2^{O(n^{\alpha})}$ ($0<\alpha<1$)} on instances of size $n$ 
  \item $\mathcal{B}$ is believed to require time $2^{\Theta(n')}$ on instances of size $n'$, i.e. it does not admit a sub-exponential time algorithm.
\end{itemize}
\vv{If there were a polynomial‐time reduction mapping each $\mathcal{B}$‐instance of size $n'$ to an $\mathcal{A}$‐instance of size growing strictly slower than $n^{1/\alpha}$, $n \sim o(n'^{1/\alpha})$, then combining this reduction with the $2^{O(n^{\alpha})}$‐time algorithm would solve $\mathcal{B}$ in time $2^{o(n')}$, contradicting ETH.  
Thus any such reduction must produce $\mathcal{A}$‐instances whose size grows at least as $n \sim \Omega(n'^{1/\alpha})$~\cite{cormen2009introduction}, essentially preserving the exponential complexity of $\mathcal{B}$.
Under ETH, no known embedding can avoid this cost, so end‐to‐end quantum algorithms for NP‐hard tasks may not outperform classical methods unless one finds embeddings with overhead lesser than $1/\alpha$, thus proving P$=$NP.
In the particular case of the MWIS problem on UD-graphs it is known that it exists a sub-exponential time algorithm running in time $2^{\mathcal{O}(\sqrt{n})}$~\cite{alber2004geometric,de2018framework}, thus leading to a quadratic overhead ($n \sim \Omega(n'^{2})$)}.
A prominent example of this is the polynomial reduction of Ref.~\cite{nguyen2023quantum} that translate arbitrary graph MWIS instances to unit‐disk graphs by means of gadgets. There, the authors observe that the size of the mapped instance could be reduced through vertex reordering. However, finding the optimal permutation of vertices requires computing the graph optimal path decomposition: an NP-hard problem in its own~\cite{robertson1986graph}. Notably, for small graphs, one can still compute an optimal path decomposition via a branching algorithm~\cite{gudmundsson2014experimental}; for larger graphs, one must resort to heuristics to find a `good enough' decomposition, which in turn raises the question of why one should not simply apply heuristic algorithms from the outset.

A similar rationale to that used for the MWIS problem in this section can be extended to arbitrary Quadratic Unconstrained Binary Optimization (QUBO) problems. When the graph structure induced by a QUBO problem exhibits favorable properties -- such as bounded treewidth, planarity, or low edge density~\cite{cazals2025identifyinghardnativeinstances} -- sub-exponential time algorithms are known to exist~\cite{fomin2019finding,koana2024subexponential}. This suggests that QUBO instances naturally embeddable in neutral atom platforms, which inherently satisfy these structural constraints, also admit sub-exponential algorithms. Consequently, any reduction aiming to transform arbitrary QUBO problems into embeddable ones will likely introduce an overhead in the number of qubits.

To conclude, let us comment that most of problems of interest for real-world applications belong to category $\mathcal{B}$: they do not admit a sub-exponential time algorithm. The ones admitting it are carefully detailed in Ref.~\cite{Woeginger2001ExactAF}. Therefore, for many concrete and practically relevant combinatorial optimization problems a significant overhead must be expected for their reductions to instances embeddable on neutral atom platforms.

\subsection{Approximation quality}
Let us now turn our attention to approximation algorithms for combinatorial optimization problems and focus, in particular, on the quality of the solution of the MWIS instances after polynomial-time reductions. Approximation algorithms offer a practical alternative for NP-hard optimization problems by trading exact optimality for provable guarantees on solution quality.
Rather than insist on a global optimum, these algorithms compute solutions that
lie within a known factor of optimality while achieving significantly improved
runtime behavior.  A premier example is the Polynomial-Time Approximation Scheme (PTAS)~\cite{vazirani2001approximation}. \vv{For any fixed $\varepsilon>0$, a PTAS is an algorithm returning in polynomial time a solution} whose cost deviates from the optimum by at most a factor $(1+\varepsilon)$ (for minimization problems). Although the dependence on $1/\varepsilon$ can be super-polynomial, the algorithm remains polynomial in the input size for each constant $\varepsilon$.
In arbitrary graphs, the MWIS  problem does not admit any $n^{1-\varepsilon}$-approximation algorithm unless P=NP~\cite{haastad1999clique}. In stark contrast, when we restrict to UD graphs with bounded density, MWIS admits both a
PTAS and an Efficient PTAS (EPTAS), with a more controlled runtime
dependence on $1/\varepsilon$~\cite{van2005approximation}. Consequently, one can compute near-optimal independent sets in polynomial time.
Any reduction that maps instances of a problem \textit{not} admitting a PTAS (e.g. MWIS on arbitrary graphs) to a problem admitting a PTAS (e.g. MWIS on UD graphs) while preserving approximation ratios would imply a PTAS for the original problem, contradicting P$\neq$ NP. 
Such a reduction would also undermine any control over solution quality: a near-optimal outcome on the transformed instance might translate back into unfeasible or far from optimal solutions on the original graph.
This fundamental loss of approximation control illustrates why reductions to geometrically simple instances cannot capture the hardness of general NP-hard problems.
In Fig.~\ref{fig:fig1}$(b)$ we show explicitly that near-optimal solutions of the MIS problem on UD graphs obtained through polynomial reductions can be arbitrarily bad solutions for the original problem.
In particular, we sample Erd\"os-Rényi graphs $G$ of size $n=22$, with $p=0.4$~\footnote{Here, $p$ is the probability of drawing an edge between two nodes, as by convention in \textit{networkx} library}. For each graph $G$ we obtain its mapping, thanks to the gadgets presented in Ref.~\cite{ebadi2022quantum}, to a graph $G'$. On $G'$ we compute all the solutions of the MIS problem up to a threshold value of the optimality gap, i.e. $\Delta(G')=0.2$, and retrieve the corresponding gap $\Delta(G)$ on the original graph $G$. \vv{The optimality gap is usually defined as the normalized distance of the cost $\mathcal{C}(\sigma)$ of a solution $\sigma$ with respect to the cost $\mathcal{C}(\sigma^*)$ of the optimal solution $\sigma^*$:
\begin{equation}\label{eq:gap}
    \Delta(\sigma)=\left| \frac{\mathcal{C}(\sigma)-\mathcal{C}(\sigma^*)}{\mathcal{C}(\sigma^*)}\right|.
\end{equation}
Here, for a given input graph $G$, we are are not interested in the specific solution $\sigma$, but rather exclusively on its optimality gap $\Delta(\sigma)$}
\vv{Therefore, we plot in orange the optimality gaps $\Delta(G)$ for the solutions $\sigma$ in $G$ as a function of the optimality gaps $\Delta(G')$ when they are used as solutions in $G'$}, averaged over 100 instances of Erd\"os-Rényi graphs. This is compared with the average optimality gap of randomly picking $m$ nodes in $G'$. We choose $m$ equal to the size of the MIS in $G$. We observe that as soon as the gap $\Delta(G')$ is larger than a threshold $\bar{\Delta}(G)'\sim 0.025$, the solutions become compatible with random sampling, witnessing that near-optimal solutions in $G'$ correspond to poor quality solution in $G$.
In the inset, we show numerically that the threshold $\bar{\Delta}(G')$ decreases with system size. We plot it as a function of the number of nodes in $G$, ranging from $n=10$ to $n=24$. Therefore, for large instances, any solution of the problem in $G'$, different from the optimal one, would be meaningless for practical purposes.

The same considerations regarding approximation guarantees extend naturally to general QUBO problems. If the original problem does not admit a PTAS, then any approximation-preserving reduction to an \vv{embeddable} QUBO instance must result in a QUBO that also does not admit a PTAS.

\section{Heuristic approach} 
As an alternative to polynomial reductions we propose here a heuristic approach.
The main idea is a recursive algorithm that will derive the MWIS solution for a general graph $G$ by constructing a valid union of sequential MWIS
solutions of embeddable subgraphs of $G'$.
We will give an overview of the different steps in the following sections.

\subsection{Greedy Lattice Subgraph mapping}

\begin{algorithm}[t]
\caption{Greedy Lattice Mapping}
\label{alg:Greedy_Lattice_Mapping}
\SetKwInOut{Input}{Input}
\SetKwInOut{Output}{Output}
\SetKwInOut{Definitions}{Definitions}

\Input{%
  Graph $G=(V_G,E_G)$,%
  Lattice $L=(V_L,E_L)$,%
  Starting node $v_s\in V_G$
}
\Output{%
  A mapping (dictionary) from a subset of $V_G$ to a subset of $V_L$
}
\Definitions{%
\begin{itemize}
  \item \textit{mapping}: dictionary $V_G \to V_L$\;
  \item \textit{isValidMapping}($g,l$): true if placing $g$ at $l$ preserves neighborhood constraints
  \end{itemize}
}

mapping $\leftarrow \{\,v_s \mapsto \text{center}(L)\,\}$\;
worklist $\leftarrow [\,v_s\,]$\;

\While{worklist is not empty}{
  $v \leftarrow$ worklist.pop()\;
  $l \leftarrow$ mapping[$v$]\;
  $U \leftarrow$ unmapped neighbors of $v$ in $G$\;
  $F \leftarrow$ unmapped neighbors of $l$ in $L$\;
  Sort $U$ by ranking strategy (e.g.\ degree)\;

  \ForEach{graph node $u\in U$}{
    \ForEach{lattice node $f\in F$}{
      \If{isValidMapping($u,f$)}{
        mapping[$u$] $\leftarrow f$\;
        worklist.append($u$)\;
        remove $f$ from $F$\;
        \textbf{break}\;
      }
    }
  }
}

\Return mapping
\end{algorithm}

\vv{
To construct an embeddable subgraph of a general input graph onto a physically realizable support lattice, we introduce a greedy algorithm which incrementally builds a valid mapping between a general graph $G$ and the nodes of a particular embeddable lattice $L$. We call this algorithm `Greedy Lattice Subgraph (GLS) mapping'. In Alg.~\ref{alg:Greedy_Lattice_Mapping} we present its pseudo-code and in App.~\ref{app:UDsubgraph_sampling} we detail a single iteration of the protocol as a toy example.
The GLS algorithm begins by choosing a single vertex in the input graph \(G\) and placing it at the central site of the target lattice \(L\). From that seed point, the algorithm always looks outward: at each step it picks one of the already‐placed graph vertices whose neighbors have not yet all been embedded, and then it considers each neighbor in turn. For each such neighbor, it inspects the free lattice sites adjacent to the already‐placed vertex and selects the first site it finds that does not break any previously established adjacency constraints — namely, whenever the neighbor has multiple already-placed graph connections, the chosen lattice location must lie adjacent on \(L\) to all of those placements. As soon as a valid site is found, that neighbor is fixed there and itself becomes eligible for further expansion. The embedding grows, marching outward from the initial node. The process terminates once every reachable graph vertex has been assigned a lattice position or no further valid placements remain, at which point the algorithm returns whatever partial mapping it has constructed.
Let us observe that the algorithm is structured such that it will always provide a valid placement for the nodes, namely respecting the connectivity of the original graph.
The candidate mappings are prioritized using a ranking — such as vertex degree and number of neighbors already assigned — that determines which unmapped graph nodes to place first.
A mapping is accepted only if it preserves the structure of the original graph: two adjacent graph nodes can only be mapped to close lattice sites, and non-adjacent graph nodes must be far away from each other on the lattice. 
In each iteration, the implementation of the algorithm removes from consideration the neighbors of the current graph node that were not compatible with any placement since it would not be possible to find valid positions further in the execution. 
The algorithm performance might depend on the target layout employed.
We explored three different calibrated layouts that could be used as a target in Pasqal Orion-$\alpha$ machine. We scored them in terms of which one was the target lattice yielding larger subgraphs after decomposition of the original graph.
We observed that for $60\%$ of the randomly sampled Érdos-Renyi graphs investigated, the largest subgraphs appeared when the target lattice was triangular. For this reason, in the following we consider a triangular lattice in our numerical simulations and experiments (See also Fig.~\ref{fig:sublattice_mapping}).}

\subsection{Recursive GLS resolution}
To solve the Maximum Weight Independent Set (MWIS) problem on a graph~\(G\), we use the GLS mapping strategy recursively. The process begins with the full graph:
\begin{equation}
  G_{1} \gets G
  \quad\text{and}\quad
  \sigma = \varnothing.
\end{equation}
Here, \(G_i\) denotes the working graph at iteration step~\(i\), and \(\sigma\) is the current partial solution—i.e., a set of vertices forming an independent set.
At each iteration~\(i\), the algorithm proceeds as follows: $(i)$ we select a subgraph \(I_i \subseteq G_i\); $(ii)$ we solve the MWIS problem on \(I_i\), and add the selected vertices to \(\sigma\); $(iii)$ we remove from \(G_i\) both the vertices in \(I_i\) and their neighborhood, defined as
  \begin{equation}
    N_{G_i}[I_i] = I_i \cup \left\{ u \in N_{G_i}(v) \mid v \in I_i \right\},
  \end{equation}
  resulting in the reduced graph \(G_{i+1}\).
This procedure is repeated until the graph is empty. The final set $\sigma$ is then a maximal independent set~\footnote{A maximal independent set is an independent set that is not a subset of any other independent set.}.

To better explore the solution space, we adopt a parallel decomposition strategy. Initially, we sample $k$ UD-subgraphs from the original graph, denoted by \(\{G^{(j)}_{2}\}_{j=1,\dots,k}\). Each subgraph is processed independently: we solve the MWIS problem and retain the top $s$ independent sets with the highest total weight, denoted by \(\sigma^{(j)}_{2}\).
At each subsequent iteration, for each current subgraph, we recursively repeat the process—generating further subgraphs, solving the MWIS problem on them, and retaining the top \(s\) partial solutions. To control complexity and avoid exponential growth, we cut the search space at every step, keeping only the best $\ell$ `branches' in a `breadth-first' fashion~\cite{Moore}, based on the solution quality.
In our implementation, we fix \(k=4\), \(s=2\), and \(\ell=10\). Eventually, the largest independent set obtained across all branches is selected as the final MWIS candidate solution.

\section{Numerical and experimental results}
\vv{
To validate our approach we perform numerical simulations and experiments for the MIS problem on Erd\"os-Rényi graphs for sizes $n$ up to 500 nodes and $p=0.5$ -- $p$ being the probability of drawing an edge between two nodes.
We use the solutions obtained with state-of-the-art classical solvers, such as `dimod' simulated annealing~\footnote{The parameters employed are the default one of the solver~\cite{dimod}. The algorithm ran on a 64-core CPU (Dual AMD Rome 7742, 128 cores total, 2.25 GHz). The time limit for resolution has been set to 2 hours and 256GB of RAM have been employed.} and `CPLEX'~ \footnote{CPLEX ran on a 64-core CPU (Dual AMD Rome 7742, 128 cores total, 2.25 GHz). As for Simulated Annealing, the time limit for resolution has been set to 2 hours and 256GB of RAM have been employed. The initial solution is set as the best solution of the Simulated Annealing} as our classical baselines.
We compare further out method -- numerically and experimentally -- with a basic classical Greedy algorithm~\footnote{A Greedy algorithm is a method that constructs a solution step by step, always selecting the option that looks best at the moment (the “greediest” choice). Once a choice is made, it is never reconsidered, which makes greedy algorithms fast and simple but means they can miss the globally optimal solution. For the Maximum Independent Set problem in consists, at each iteration, in $(i)$ selecting a vertex (often the one of minimum degree in the current subgraph), $(ii)$ adding adding the vertex to the independent set, $(iii)$ removing the chosen vertex and all its neighbors from further consideration, $(iv)$ until no vertices remain. This process runs in near-linear time and produces a maximal independent set.} in order to provide evidence of speed-up with respect to a naive approach.
On the quantum mechanical side, we solve the subgraphs generated by our algorithm by quantum annealing. We drive our neutral atoms system adiabatically towards the ground state of the Rydberg Hamiltonian that natively encodes the MWIS problem.
We perform both numerical simulations of the neutral atoms QPU and experiments with the Orion-$\alpha$ machine.
In the following, first we describe the experimental platforms and detail the parameters of the numerical simulations, then we present the results.}

\begin{figure}
    \centering
\includegraphics[width=\linewidth]{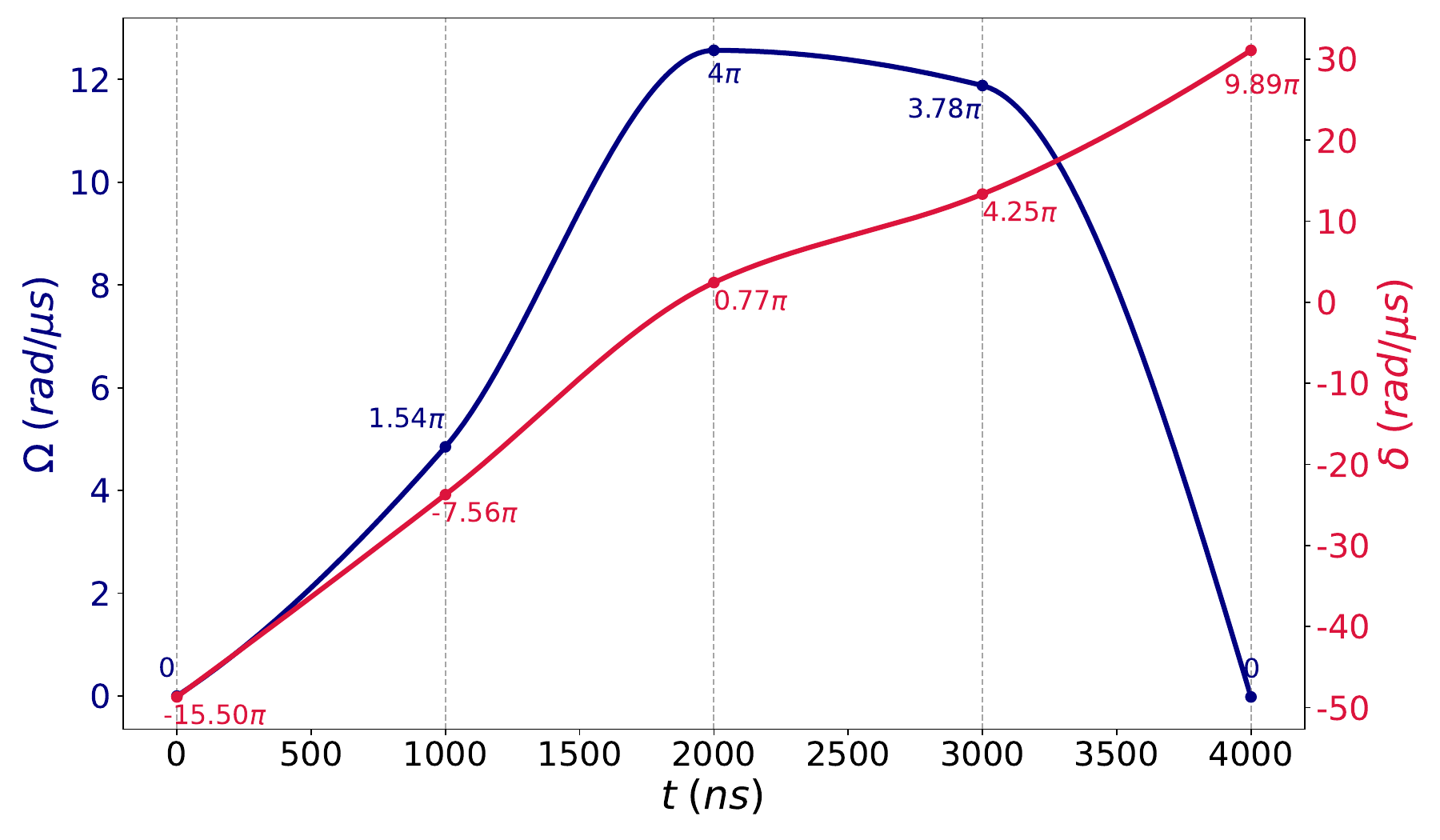}
    \caption{Annealing protocol of time dependent controls $\Omega$ and $\delta$ employed to find the MIS of the graphs. The values and the positions of the interpolated points are written on the figure. We report them here for convenience: $(t,\Omega,\delta)\in
    $$\{(0,0,-15.50\pi)$, $(1000,1.54\pi,-7.56\pi)$, $(2000,4\pi,0.77\pi)$, $(3000,3.78\pi,4.25\pi)$, $(4000,0,9.89\pi)\}$}
    \label{fig:annealing-protocol}
\end{figure}
\vv{
We employ a neutral-atom based QPU made of single $^{87}$Rb atoms trapped in arrays of optical tweezers~\cite{barredo2018synthetic,Nogrette14,browaeys2020many,henriet2020quantum,Morgado2021}. Qubits are encoded onto the ground state $\ket{0}=\ket{5S_{1/2},F=2,m_F=2}$ and the Rydberg state $\ket{1}=\ket{60S_{1/2}, m_J=1/2}$ of the atoms.
The dynamics of a collection of $N$ qubits at positions $\boldsymbol{r}$ are governed by the following Hamiltonian $\hat H(t)=\hat H_{\rm drive}(t)+\hat H_{\rm cost}(t)$, where:
\begin{equation}
\hat{H}(t)=\frac{\Omega(t)}{2}\sum_{i=1}^N\hat\sigma_i^x+\delta(t)\sum_{i=1}^N\epsilon_i\hat n_i+\sum_{i<j}U(|r_i-r_j|)\hat n_i \hat n_j.
\end{equation}
where we set $\hbar=1$.
Here, $\hat{\sigma}_i^\alpha$
are Pauli matrices and $\hat n_i=\ket{1}_i\bra{1}_i$. The two time-dependent control fields are the Rabi frequency $\Omega(t)$ and the detuning $\delta(t)$. The term $U(|r_i-r_j|)=C_6/|r_i-r_j|^{6}$ is the position-dependent interaction, with $C_6/h=138~$GHz$\cdot \mu {\rm m}^6$ for the Rydberg state considered \cite{Beguin_2013,ibali2017}. 
For a graph $\Graph$, the MWIS cost function can be encoded on the Rydberg Hamiltonian according to the following equation:
\begin{equation}  C_\Graph(\hat n)=\hat C_\Graph=-\sum_{i\in\vertices}w_i\hat n_i+\alpha \sum_{i<j}\hat n_i \hat n_j.
\end{equation}
This is equivalent to the MWIS formulation in Fig.~\ref{fig:fig1}, where the binary variable $x_i$ is substituted by the number operator $n_i$.
The nodes are weighted with $w_i\in[0,1]$, and edges between adjacent nodes carry a uniform weight of $\alpha>0$. In the case of the weights $w_i$ being uniformly $1$, the problem to solve becomes an MIS. The ground state of $\hat C_\Graph$ manifests as a coherent superposition of MWIS classical product states. Such an operator finds correspondence with the Hamiltonian $H$ (with $\delta>0$) of $N$ atoms well-positioned such that:
\begin{equation}
        \hat{H} \approx \delta\left(-\sum_{i=1}^N w_i\hat n_i+\frac{U(r_b)}{\delta} \sum_{i<j}\hat n_i \hat n_j\right) =\delta \hat C_\Graph
\end{equation}
where we take $\epsilon_i=w_i$ and the ratio between the nearest-neighbors interaction $U(r_b)$ and the detuning encodes the uniform edge weight $\alpha$. 
By preparing the ground state of $\hat H_{\rm cost}$, and subsequently sampling it $n_{\rm shots}$ times, it becomes possible to measure not just one, but all the MWISs of $\Graph$ as they are degenerate.
The procedure employed for solving the MWIS problem is the following.
A given graph is generated. The quantum state is initialized as $\ket{\psi}=\ket{0}^{\otimes N}$ and driven towards the correct final target state through a slow evolution according to the paradigm of adiabatic quantum computation~\cite{farhi2014quantum}.
The particular time-dependence of the control field is called annealing schedule.
In this work, we have adopted the one utilized shown in Fig.~\ref{fig:annealing-protocol} and Ref.~\cite{cazals2025identifyinghardnativeinstances}. 
The initial and final values of $\Omega$ have been fixed such that the initial Hamiltonian ground state corresponds to the prepared initial state, and the final Hamiltonian encodes the optimization problem.
The initial detuning value instead has been fixed to the minimum possible value, compatible with hardware capabilities.
All the other points have been optimized in Ref.~\cite{cazals2025identifyinghardnativeinstances}, to maximize the probability of finding the MIS, over a set of random unit-disk graphs on a triangular layout with lattice spacing $a=5\mu m$ (target of the GLS algorithm), with $N=20$ and densities varying from 0.1 to 1 in steps of 0.1.
Arranging the atoms and running the optimised sequence, allows preparing the quantum system in a state that encodes the solution of the MWIS problem on the particular lattice.
Sampling from that quantum superposition returns a distribution of bitstrings which either constitute perfect (MWIS) or approximated (MWIS-$k$) solutions for the problem.
We performed numerical simulations of the time evolution by evolving the quantum state expressed as a Matrix Product State (MPS) according to the Time-dependent Variational Principle (TDVP)~\cite{Bidzhiev_pasqal_emulators_2025}.
The labeling of the atoms for the MPS unraveling on the triangular lattice has been optimized by minimizing the bandwidth of the Hamiltonian MPO~\cite{Bidzhiev_pasqal_emulators_2025}.
The maximum allowed bond dimension of the MPS has been fixed to $\chi=400$, which implies a truncation error $\sim 10^{-7}$.
The time-step for the TDVP has been set to $dt= 0.002$.
At the end of the evolution, we sample $n_{\text{shots}}=1000$ bitstrings, which ensure us to find, at least once, the MIS of the graph for sizes comparable to the ones of the subgraphs generated by the GLS protocol.
While the QPU is subject to noise we have not applied any error mitigation routine in this work nor we have performed noisy simulations.
The rationale behind this is that the GLS approach is inherently robust against noise as it only relies on a few optimal bitstrings to be sampled. Therefore, one can neglect the noise unless it is strong enough to prevent the protocol to find those solutions. We claim that this is not the case by considering as a reference point, the experimental results of Ref.~\cite{cazals2025identifyinghardnativeinstances}.}

\vv{We show the results of numerical simulations and experiments in comparison with classical approaches in Fig.~\ref{fig:fig1}$(c)$, where we plot the optimality gap as in Eq.~\ref{eq:gap}. We present additional numerical results in App.~\ref{sec:additional_results}.}
For completeness, in the inset of Fig.~\ref{fig:fig1}$(c)$ we plot the occurrences of the sizes of the largest subgraphs found: obtaining large subgraphs is essential for faster resolution, since an excess of small subgraphs would become a significant bottleneck.
Overall, the results in Fig.~\ref{fig:fig1}$(c)$ show that we are able to solve the MIS problem with better performances with respect to a Greedy approach and for a number of nodes that goes beyond today's experimental capabilities of neutral atoms platforms -- by decomposing the original problem in medium sized subgraphs that can be handled by the QPU. QPU results, shown in red, are indeed in agreement with numerical simulations.

\section{Discussion and outlook} 
In this work, we have examined the limits of polynomial reductions for mapping general‐graph MWIS instances onto unit‐disk architectures. Leveraging the Exponential Time Hypothesis, we proved that any reduction from a problem lacking subexponential‐time algorithms to one admitting such algorithms must incur a quadratic blow-up in instance size, thus potentially erasing any quantum speedup. We further demonstrated that near‐optimal solutions on the reduced unit-disk instances can correspond to arbitrarily poor solutions on the original graphs, undermining any control over approximation quality.

Motivated by these findings, we introduced a heuristic that avoids the reduction quadratic overhead by iteratively extracting UD‐embeddable subgraphs, solving each one on precalibrated atomic layouts, and merging their independent sets into a global solution. Numerical simulations on Erdős–Rényi graphs (up to $n=500$) and preliminary experimental results confirm that our approach surpasses classical Greedy algorithms and extends the practical reach of neutral‐atom MWIS solvers beyond current device limitations.

We observe that in this work we have focused on polynomial reductions that aim at solving the full problem within QPU capabilities. As an alternative to gadget‐based approaches, one could pursue hybrid embedding schemes that leverage classical optimization together with quantum resources. Often, in such approaches, the original optimization problem is decomposed in several building blocks which are solved using the most suited classical or quantum routine. Such hybrid schemes have been proved useful in several instances such as vertex-coloring problems and Mixed-Integer Linear Programming~\cite{wesley,yassinebenders,dalyac2024graph,lucasleclerc2023}.

Furthermore, a possible direction to extend the range of neutral atoms today's capabilities could be Hamiltonian engineering, where neutral atoms can be programmed to realize complex, non-local interactions. Recent work has demonstrated the feasibility of tuning a diverse range of interactions and connectivity opening up new possibilities for quantum optimization that classical solvers struggle to address, even though presenting serious limitations in terms of robustness to external noise, experimental imperfections, and short coherence time~\cite{scholl2022microwave,barredo2018synthetic,lee2016three,votto2024universal,parra2020digital,garcia2024digital,goldman2014periodically,rajabi2019dynamical}.

\section*{Acknowledgements}
We thank the Hardware and Cloud team from Pasqal.
We are grateful to the Emulators team of Pasqal for providing the state vector and tensor network emulators used for this work~\cite{Bidzhiev_pasqal_emulators_2025}.
We thank Louis-Paul Henry, Malory Marin, Florian Sikora and Rémy Watrigant for useful feedback.
VV is grateful to Yassine Naghmouchi for helpful discussions about the classical optimization problems.

\section*{Authors contribution}
PC, CD and WC conceived the main idea of the manuscript.
AS and PC devised and implemented the heuristic algorithm. AS, PC, VO and VV performed the numerical simulations and the experiments via Pasqal's cloud platform. All the authors discussed the results. VV wrote the manuscript with the help of all the authors.

\bibliography{biblio}

\onecolumngrid
\appendix
\appendixpage

\section{Asymptotic notation}\label{sec:asymptotic_notation}
Asymptotic notation provides a language for describing the growth rates of functions, particularly the running time or space usage of algorithms, as the input size \(n\) becomes large.  By suppressing constant factors and lower‐order terms, these notations focus on the dominant behavior and allow meaningful comparisons between algorithms.
Here we summarize the notation for the ease of the reader~\cite{cormen2009introduction}.\\
\begin{description}
\item[Big-\(O\) (Upper Bound)] We say
\[
  f(n) = O\bigl(g(n)\bigr)
\]
if there exist constants \(c > 0\) and \(n_0\) such that
\[
  0 \;\le\; f(n) \;\le\; c\,g(n)
  \quad\text{for all }n\ge n_0.
\]
Thus \(f(n)\) does not grow faster than \(g(n)\) up to constant factors.\\

\item[Big-\(\Omega\) (Lower Bound)] We say
\[
  f(n) = \Omega\bigl(g(n)\bigr)
\]
if there exist constants \(c > 0\) and \(n_0\) such that
\[
  0 \;\le\; c\,g(n) \;\le\; f(n)
  \quad\text{for all }n\ge n_0.
\]
Therefore \(f(n)\) grows at least as fast as \(g(n)\) up to constant factors.\\
\item[Big-\(\Theta\) (Tight Bound)] We say
\[
  f(n) = \Theta\bigl(g(n)\bigr)
\]
if there exist constants \(c_1, c_2 > 0\) and \(n_0\) such that
\[
  c_1\,g(n) \;\le\; f(n) \;\le\; c_2\,g(n)
  \quad\text{for all }n\ge n_0.
\]
Equivalently, \(f(n)\) is both \(O(g(n))\) and \(\Omega(g(n))\).  This denotes the exact asymptotic growth rate, up to constant factors.\\

\item[Little-\(o\) (Strictly Smaller)] We say
\[
  f(n) = o\bigl(g(n)\bigr)
\]
if for every \(\varepsilon>0\) there exists \(n_0\) such that
\[
  0 \;\le\; f(n) \;<\;\varepsilon\,g(n)
  \quad\text{for all }n\ge n_0,
\]
or equivalently \(\lim_{n\to\infty} f(n)/g(n) = 0\).  This indicates \(f\) grows strictly slower than \(g\).

\end{description}

\section{Greedy Subgraph Sampling}\label{app:UDsubgraph_sampling}

In Fig.~\ref{fig:iteration} we detail a single iteration of the Greedy Subgraph Sampling algorithm described in the main text. We want to map the graph $G$ onto the triangular lattice $L$.
\begin{description}
    \item[Step 0] We pick a node in $L$ and take into account its neighborhood, i.e. $\{l_1,l_2,l_3,l_4\}$.
    Then, we randomly pick a node in $G$, denoted in red, and we consider the neighborhoods of the current node in graph $G$, i.e. $\{n_1,n_2,n_3\}$.
    \item[Step 1] The first move is always accepted as there are no constraints. For instance, we set $\{ n_{1} \rightarrow l_{4} \}$ as a valid mapping,  we add it to the subgraph mapping. For ease of visualization, we color the mapped nodes.
    \item[Step 2] Let us now consider $n_{2}$ as neighbor of the current graph node and $l_{3}$ as neighbor of the current lattice node (Step 2 of Fig.~\ref{fig:iteration}). We know that node $n_{1}$ was mapped to $l_{4}$. If we map $n_{2}$ to $l_{3}$, we obtain a valid assignment.
    \item[Step 3] Then let us consider $n_{3}$ as neighbor of the current graph node and $l_{2}$ as neighbor of the current lattice node. In this setting, the edge connecting $n_{1}$ to $n_{3}$ would be neglected since there is no edge connecting $l_{2}$ to $l_{4}$ (denoted with a dashed red line). Therefore, $\{ n_{3} \rightarrow l_{2} \}$ is not a valid mapping.
    \item[Step 4] Then, we consider $n_{3}$ as neighbor of the current graph node and $l_{1}$ as neighbor of the current lattice node. It is not a valid mapping again because the edges connecting $n_{3}$ to $n_{1}$ and $n{2}$ would be neglected (denoted with dashed red lines).
    \item[Stop] There are no possible assignments. The algorithm provides the following partial mapping:
    \begin{equation}
        \{n_0,n_1,n_2\}\rightarrow \{l_0, l_4, l_3\}.
    \end{equation}
\end{description}

\tikzset{create hullnodes/.code={
        \global\edef\namelist{#1}
        \foreach [count=\counter] \nodename in \namelist {
            \global\edef\numberofnodes{\counter}
            \node at (\nodename) [draw=none,name=hullnode\counter] {};
        }
        \node at (hullnode\numberofnodes) [name=hullnode0,draw=none] {};
        \pgfmathtruncatemacro\lastnumber{\numberofnodes+1}
        \node at (hullnode1) [name=hullnode\lastnumber,draw=none] {};
    },mynode/.style={outer sep=0pt,draw,shape=circle,minimum size=11mm,inner
sep=0pt}}

\begin{figure}
\begin{tikzpicture}
    \hspace{4ex}
    \node[draw, circle, inner sep=2pt, color=blue] (a) at (0,0) {};
    \node[draw, circle, inner sep=2pt, color=red, fill] (b) at (2,0) {};
    \node[draw, circle, inner sep=2pt, color=blue] (c) at (2,2) {};
    \node[draw, circle, inner sep=2pt, color=blue] (d) at (0,2) {};
    \node[draw, circle, inner sep=2pt] (e) at (1,3) {};
    
    \draw (a) -- (b) -- (c) -- (d) -- (a);
    \draw (c) -- (e) -- (d);
    \draw (d) -- (b);
    \draw (a) -- (c);
    
    \node[below left = 0.2cm and 0.2cm] at (a) {\textcolor{blue}{$n_{3}$}};
    \node[below right, align=center, color=blue] at (b) {$n_0$ 
 \textcolor{red}{current} \\   \textcolor{red}{graph node}};
    \node[above right = 0.2cm and 0.2cm] at (c) {\textcolor{blue}{$n_{1}$}};
    \node[above left = 0.2cm and 0.2cm] at (d) {\textcolor{blue}{$n_{2}$}};
    \def\side{2} 
    \def\height{1.73} 

    \foreach \row in {0,1} {
        \foreach \col in {0,1} {
            \coordinate (A) at (5 + \col*\side + \row*\side/2, \row*\height);
            \coordinate (B) at (5 + \col*\side + \side + \row*\side/2, \row*\height);
            \coordinate (C) at (5 + \col*\side + \side/2 + \row*\side/2, \row*\height + \height);


            \node[draw, circle, inner sep=2pt] (a_\row_\col) at (A) {};
            \node[draw, circle, inner sep=2pt] (b_\row_\col) at (B) {};
            \node[draw, circle, inner sep=2pt] (c_\row_\col) at (C) {};

            \draw (a_\row_\col) -- (b_\row_\col);
            \draw (c_\row_\col) -- (b_\row_\col);
            \draw (a_\row_\col) -- (c_\row_\col);
            
        }
    }

    \draw (c_1_1) -- (c_1_0);
    \draw (b_0_1) -- (b_1_1);

    \node[draw, circle, inner sep=2pt, color=red, fill] at (a_0_1) {};
    \node[below, align=center, color=green] at (a_0_1) {$l_0$   
\textcolor{red}{current} \\ \textcolor{red}{lattice node}};

    \node[draw, circle, inner sep=2pt, color=green] at (a_1_0) {};
    \node[above left = 0.3cm and 0.3cm, align=center] at (a_1_0) {\textcolor{green}{$l_{2}$}};

    \node[draw, circle, inner sep=2pt, color=green] at (a_1_1) {};
    \node[above right = 0.3cm and 0.5cm, align=center] at (a_1_1) {\textcolor{green}{$l_{3}$}};
    
    \node[draw, circle, inner sep=2pt, color=green] at (a_0_0) {};
    \node[below = 0.4cm, align=center] at (a_0_0) {\textcolor{green}{$l_{1}$}};

    \node[draw, circle, inner sep=2pt, color=green] at (b_0_1) {};
    \node[below = 0.4cm, align=center] at (b_0_1) {\textcolor{green}{$l_{4}$}};

    \node[above = 1cm, draw=none, fill=none] (inv) at (a_0_1) {};

    \node[above right = 0.25cm and 0.85cm, align=center] at (b_0_1) {\textcolor{green}{$N_{L}(\text{current}$} \\ \textcolor{green}{$\text{lattice node})$}};
    
    \node[above left = 2.9cm and 1 cm] at (a) {Graph $G$};
    \node[left = 1.5cm] at (a) {\underline{Step 0}};
    \node[above right = 3cm and 3 cm] at (a_0_1) {Lattice $L$};

\end{tikzpicture}

\vspace{2ex}

\begin{tikzpicture}
    \node[draw, circle, inner sep=2pt, color=blue] (a) at (0,0) {};
    \node[draw, circle, inner sep=2pt, color=red, fill] (b) at (2,0) {};
    \node[draw, circle, inner sep=2pt, color=blue, fill] (c) at (2,2) {};
    \node[draw, circle, inner sep=2pt, color=blue] (d) at (0,2) {};
    \node[draw, circle, inner sep=2pt] (e) at (1,3) {};
    
    \draw (a) -- (b) -- (c) -- (d) -- (a);
    \draw (c) -- (e) -- (d);
    \draw (d) -- (b);
    \draw (a) -- (c);
    \draw[line width=2pt] (b) -- (c);
    
    \node[below left = 0.2cm and 0.2cm] at (a) {\textcolor{blue}{$n_{3}$}};
    \node[below right= 0.2cm and 0.2cm, align=center, color=blue] at (b) {$n_0$};
    \node[above right = 0.2cm and 0.2cm] at (c) {\textcolor{blue}{$n_{1}$}};
    \node[above left = 0.2cm and 0.2cm] at (d) {\textcolor{blue}{$n_{2}$}};
    \node[below left = 0.75cm and 1cm] at (d) {\textcolor{white}{$N_{G}(\text{current node})$}};
    \node[draw, fit=(c), inner sep=2pt, line width=1pt, color=blue] {};

    \def\side{2} 
    \def\height{1.73} 

    \foreach \row in {0,1} {
        \foreach \col in {0,1} {
            \coordinate (A) at (5 + \col*\side + \row*\side/2, \row*\height);
            \coordinate (B) at (5 + \col*\side + \side + \row*\side/2, \row*\height);
            \coordinate (C) at (5 + \col*\side + \side/2 + \row*\side/2, \row*\height + \height);


            \node[draw, circle, inner sep=2pt] (a_\row_\col) at (A) {};
            \node[draw, circle, inner sep=2pt] (b_\row_\col) at (B) {};
            \node[draw, circle, inner sep=2pt] (c_\row_\col) at (C) {};

            \draw (a_\row_\col) -- (b_\row_\col);
            \draw (c_\row_\col) -- (b_\row_\col);
            \draw (a_\row_\col) -- (c_\row_\col);
            
        }
    }

    \draw (c_1_1) -- (c_1_0);
    \draw (b_0_1) -- (b_1_1);

    \node[draw, circle, inner sep=2pt, color=red, fill] at (a_0_1) {};
    \node[below = 0.4cm, align=center, color=green] at (a_0_1) {$l_0$};

    \node[draw, circle, inner sep=2pt, color=green] at (a_1_0) {};
    \node[above left = 0.3cm and 0.3cm, align=center] at (a_1_0) {\textcolor{green}{$l_{2}$}};

    \node[draw, circle, inner sep=2pt, color=green] at (a_1_1) {};
    \node[above right = 0.3cm and 0.5cm, align=center] at (a_1_1) {\textcolor{green}{$l_{3}$}};
    
    \node[draw, circle, inner sep=2pt, color=green] at (a_0_0) {};
    \node[below = 0.4cm, align=center] at (a_0_0) {\textcolor{green}{$l_{1}$}};

    \node[draw, circle, inner sep=2pt, color=green, fill] at (b_0_1) {};
    \node[below = 0.4cm, align=center] at (b_0_1) {\textcolor{green}{$l_{4}$}};
    \node[draw, fit=(b_0_1), inner sep=2pt, line width=1pt, color=green] {};

    \node[above = 1cm, draw=none, fill=none] (inv) at (a_0_1) {};

    \node[above right = 0.25cm and 0.85cm, align=center] at (b_0_1) {\textcolor{white}{$N_{L}(\text{current}$} \\ \textcolor{white}{$\text{lattice node})$}};
    
    \draw[line width=2pt] (a_0_1) -- (b_0_1);
    \node[left = 1.5cm] at (a) {\underline{Step 1}};
\end{tikzpicture}

\vspace{2ex}

\begin{tikzpicture}
    \node[draw, circle, inner sep=2pt, color=blue] (a) at (0,0) {};
    \node[draw, circle, inner sep=2pt, color=red, fill] (b) at (2,0) {};
    \node[draw, circle, inner sep=2pt, color=blue, fill] (c) at (2,2) {};
    \node[draw, circle, inner sep=2pt, color=blue, fill] (d) at (0,2) {};
    \node[draw, circle, inner sep=2pt] (e) at (1,3) {};
    
    \draw (a) -- (b) -- (c) -- (d) -- (a);
    \draw (c) -- (e) -- (d);
    \draw (d) -- (b);
    \draw (a) -- (c);
    \draw[line width=2pt] (b) -- (c);
    \draw[line width=2pt] (d) -- (c);
    \draw[line width=2pt] (d) -- (b);
    
    \node[below left = 0.2cm and 0.2cm] at (a) {\textcolor{blue}{$n_{3}$}};
    \node[below right= 0.2cm and 0.2cm, align=center, color=blue] at (b) {$n_0$};
    \node[above right = 0.2cm and 0.2cm] at (c) {\textcolor{blue}{$n_{1}$}};
    \node[above left = 0.2cm and 0.2cm] at (d) {\textcolor{blue}{$n_{2}$}};
    \node[below left = 0.75cm and 1cm] at (d) {\textcolor{white}{$N_{G}(\text{current node})$}};
    \node[draw, fit=(d), inner sep=2pt, line width=1pt, color=blue] {};

    \def\side{2} 
    \def\height{1.73} 

    \foreach \row in {0,1} {
        \foreach \col in {0,1} {
            \coordinate (A) at (5 + \col*\side + \row*\side/2, \row*\height);
            \coordinate (B) at (5 + \col*\side + \side + \row*\side/2, \row*\height);
            \coordinate (C) at (5 + \col*\side + \side/2 + \row*\side/2, \row*\height + \height);


            \node[draw, circle, inner sep=2pt] (a_\row_\col) at (A) {};
            \node[draw, circle, inner sep=2pt] (b_\row_\col) at (B) {};
            \node[draw, circle, inner sep=2pt] (c_\row_\col) at (C) {};

            \draw (a_\row_\col) -- (b_\row_\col);
            \draw (c_\row_\col) -- (b_\row_\col);
            \draw (a_\row_\col) -- (c_\row_\col);
            
        }
    }

    \draw (c_1_1) -- (c_1_0);
    \draw (b_0_1) -- (b_1_1);

    \node[draw, circle, inner sep=2pt, color=red, fill] at (a_0_1) {};
    \node[below= 0.4cm, align=center,color=green] at (a_0_1) {$l_0$};

    \node[draw, circle, inner sep=2pt, color=green] at (a_1_0) {};
    \node[above left = 0.3cm and 0.3cm, align=center] at (a_1_0) {\textcolor{green}{$l_{2}$}};

    \node[draw, circle, inner sep=2pt, color=green, fill] at (a_1_1) {};
    \node[above right = 0.3cm and 0.5cm, align=center] at (a_1_1) {\textcolor{green}{$l_{3}$}};
    
    \node[draw, circle, inner sep=2pt, color=green] at (a_0_0) {};
    \node[below = 0.4cm, align=center] at (a_0_0) {\textcolor{green}{$l_{1}$}};

    \node[draw, circle, inner sep=2pt, color=green, fill] at (b_0_1) {};
    \node[below = 0.4cm, align=center] at (b_0_1) {\textcolor{green}{$l_{4}$}};
    \node[draw, fit=(a_1_1), inner sep=2pt, line width=1pt, color=green] {};

    \node[above = 1cm, draw=none, fill=none] (inv) at (a_0_1) {};

    \node[above right = 0.25cm and 0.85cm, align=center] at (b_0_1) {\textcolor{white}{$N_{L}(\text{current}$} \\ \textcolor{white}{$\text{lattice node})$}};

    \draw[line width=2pt] (a_0_1) -- (b_0_1);
    \draw[line width=2pt] (a_0_1) -- (a_1_1);
    \draw[line width=2pt] (a_1_1) -- (b_0_1);
    \node[left = 1.5cm] at (a) {\underline{Step 2}};
\end{tikzpicture}

\vspace{2ex}

\begin{tikzpicture}
    \node[draw, circle, inner sep=2pt, color=blue] (a) at (0,0) {};
    \node[draw, circle, inner sep=2pt, color=red, fill] (b) at (2,0) {};
    \node[draw, circle, inner sep=2pt, color=blue, fill] (c) at (2,2) {};
    \node[draw, circle, inner sep=2pt, color=blue, fill] (d) at (0,2) {};
    \node[draw, circle, inner sep=2pt] (e) at (1,3) {};
    
    \draw (a) -- (b) -- (c) -- (d) -- (a);
    \draw (c) -- (e) -- (d);
    \draw (d) -- (b);
    \draw (a) -- (c);
    \draw[line width=2pt] (b) -- (c);
    \draw[line width=2pt] (d) -- (c);
    \draw[line width=2pt] (d) -- (b);
    
    \node[below left = 0.2cm and 0.2cm] at (a) {\textcolor{blue}{$n_{3}$}};
    \node[below right= 0.2cm and 0.2cm, align=center,color=blue] at (b) {$n_0$};
    \node[above right = 0.2cm and 0.2cm] at (c) {\textcolor{blue}{$n_{1}$}};
    \node[above left = 0.2cm and 0.2cm] at (d) {\textcolor{blue}{$n_{2}$}};
    \node[below left = 0.75cm and 1cm] at (d) {\textcolor{white}{$N_{G}(\text{current node})$}};
    \node[draw, fit=(a), inner sep=2pt, line width=1pt, color=blue] {};

    \def\side{2} 
    \def\height{1.73} 

    \foreach \row in {0,1} {
        \foreach \col in {0,1} {
            \coordinate (A) at (5 + \col*\side + \row*\side/2, \row*\height);
            \coordinate (B) at (5 + \col*\side + \side + \row*\side/2, \row*\height);
            \coordinate (C) at (5 + \col*\side + \side/2 + \row*\side/2, \row*\height + \height);


            \node[draw, circle, inner sep=2pt] (a_\row_\col) at (A) {};
            \node[draw, circle, inner sep=2pt] (b_\row_\col) at (B) {};
            \node[draw, circle, inner sep=2pt] (c_\row_\col) at (C) {};

            \draw (a_\row_\col) -- (b_\row_\col);
            \draw (c_\row_\col) -- (b_\row_\col);
            \draw (a_\row_\col) -- (c_\row_\col);
            
        }
    }

    \draw (c_1_1) -- (c_1_0);
    \draw (b_0_1) -- (b_1_1);

    \node[draw, circle, inner sep=2pt, color=red, fill] at (a_0_1) {};
    \node[below=0.4cm, align=center, color=green] at (a_0_1) {$l_0$};

    \node[draw, circle, inner sep=2pt, color=green] at (a_1_0) {};
    \node[above left = 0.3cm and 0.3cm, align=center] at (a_1_0) {\textcolor{green}{$l_{2}$}};

    \node[draw, circle, inner sep=2pt, color=green, fill] at (a_1_1) {};
    \node[above right = 0.3cm and 0.5cm, align=center] at (a_1_1) {\textcolor{green}{$l_{3}$}};
    
    \node[draw, circle, inner sep=2pt, color=green] at (a_0_0) {};
    \node[below = 0.4cm, align=center] at (a_0_0) {\textcolor{green}{$l_{1}$}};

    \node[draw, circle, inner sep=2pt, color=green, fill] at (b_0_1) {};
    \node[below = 0.4cm, align=center] at (b_0_1) {\textcolor{green}{$l_{4}$}};
    \node[draw, fit=(a_1_0), inner sep=2pt, line width=1pt, color=green] {};

    \node[above = 1cm, draw=none, fill=none] (inv) at (a_0_1) {};

    \node[above right = 0.25cm and 0.85cm, align=center] at (b_0_1) {\textcolor{white}{$N_{L}(\text{current}$} \\ \textcolor{white}{$\text{lattice node})$}};

    \draw[line width=2pt] (a_0_1) -- (b_0_1);
    \draw[line width=2pt] (a_0_1) -- (a_1_1);
    \draw[line width=2pt] (a_1_1) -- (b_0_1);
    \draw[dashed, red] (b_0_1) -- (a_1_0);
    \node[left = 1.5cm] at (a) {\underline{Step 3}};
\end{tikzpicture}

\vspace{2ex}
\begin{tikzpicture}
    \node[draw, circle, inner sep=2pt, color=blue] (a) at (0,0) {};
    \node[draw, circle, inner sep=2pt, color=red, fill] (b) at (2,0) {};
    \node[draw, circle, inner sep=2pt, color=blue, fill] (c) at (2,2) {};
    \node[draw, circle, inner sep=2pt, color=blue, fill] (d) at (0,2) {};
    \node[draw, circle, inner sep=2pt] (e) at (1,3) {};
    
    \draw (a) -- (b) -- (c) -- (d) -- (a);
    \draw (c) -- (e) -- (d);
    \draw (d) -- (b);
    \draw (a) -- (c);
    \draw[line width=2pt] (b) -- (c);
    \draw[line width=2pt] (d) -- (c);
    \draw[line width=2pt] (d) -- (b);
    
    \node[below left = 0.2cm and 0.2cm] at (a) {\textcolor{blue}{$n_{3}$}};
    \node[below right= 0.2cm and 0.2cm, align=center,color=blue] at (b) {$n_0$};
    \node[above right = 0.2cm and 0.2cm] at (c) {\textcolor{blue}{$n_{1}$}};
    \node[above left = 0.2cm and 0.2cm] at (d) {\textcolor{blue}{$n_{2}$}};
    \node[below left = 0.75cm and 1cm] at (d) {\textcolor{white}{$N_{G}(\text{current node})$}};
    \node[draw, fit=(a), inner sep=2pt, line width=1pt, color=blue] {};

    \def\side{2} 
    \def\height{1.73} 

    \foreach \row in {0,1} {
        \foreach \col in {0,1} {
            \coordinate (A) at (5 + \col*\side + \row*\side/2, \row*\height);
            \coordinate (B) at (5 + \col*\side + \side + \row*\side/2, \row*\height);
            \coordinate (C) at (5 + \col*\side + \side/2 + \row*\side/2, \row*\height + \height);


            \node[draw, circle, inner sep=2pt] (a_\row_\col) at (A) {};
            \node[draw, circle, inner sep=2pt] (b_\row_\col) at (B) {};
            \node[draw, circle, inner sep=2pt] (c_\row_\col) at (C) {};

            \draw (a_\row_\col) -- (b_\row_\col);
            \draw (c_\row_\col) -- (b_\row_\col);
            \draw (a_\row_\col) -- (c_\row_\col);
            
        }
    }

    \draw (c_1_1) -- (c_1_0);
    \draw (b_0_1) -- (b_1_1);

    \node[draw, circle, inner sep=2pt, color=red, fill] at (a_0_1) {};
    \node[below=0.4cm, align=center, color=green] at (a_0_1) {$l_0$};

    \node[draw, circle, inner sep=2pt, color=green] at (a_1_0) {};
    \node[above left = 0.3cm and 0.3cm, align=center] at (a_1_0) {\textcolor{green}{$l_{2}$}};

    \node[draw, circle, inner sep=2pt, color=green, fill] at (a_1_1) {};
    \node[above right = 0.3cm and 0.5cm, align=center] at (a_1_1) {\textcolor{green}{$l_{3}$}};
    
    \node[draw, circle, inner sep=2pt, color=green] at (a_0_0) {};
    \node[below = 0.4cm, align=center] at (a_0_0) {\textcolor{green}{$l_{1}$}};
    \node[draw, circle, inner sep=2pt, color=green, fill] at (b_0_1) {};
    \node[below = 0.4cm, align=center] at (b_0_1) {\textcolor{green}{$l_{4}$}};
    \node[draw, fit=(a_0_0), inner sep=2pt, line width=1pt, color=green] {};
    \node[above = 1cm, draw=none, fill=none] (inv) at (a_0_1) {};
    \node[above right = 0.25cm and 0.85cm, align=center] at (b_0_1) {\textcolor{white}{$N_{L}(\text{current}$} \\ \textcolor{white}{$\text{lattice node})$}};
    \draw[line width=2pt] (a_0_1) -- (b_0_1);
    \draw[line width=2pt] (a_0_1) -- (a_1_1);
    \draw[line width=2pt] (a_1_1) -- (b_0_1);
    \draw[dashed, red, bend right=20] (b_0_1) to (a_0_0);
    \draw[dashed, red] (b_1_0) to (a_0_0);
    \node[left = 1.5cm] at (a) {\underline{Step 4}};
\end{tikzpicture}
\caption{Single iteration of the Greedy Lattice Subgraph mapping. We pick a random node $n_0$ in the original graph $G$ and $l_0$ in the lattice $L$. We grow the mapping exploring the neighborhood of the selected node, till no more assignments are possible.}
\label{fig:iteration}
\end{figure}

\begin{figure}
    \centering    \includegraphics[width=0.7\linewidth]{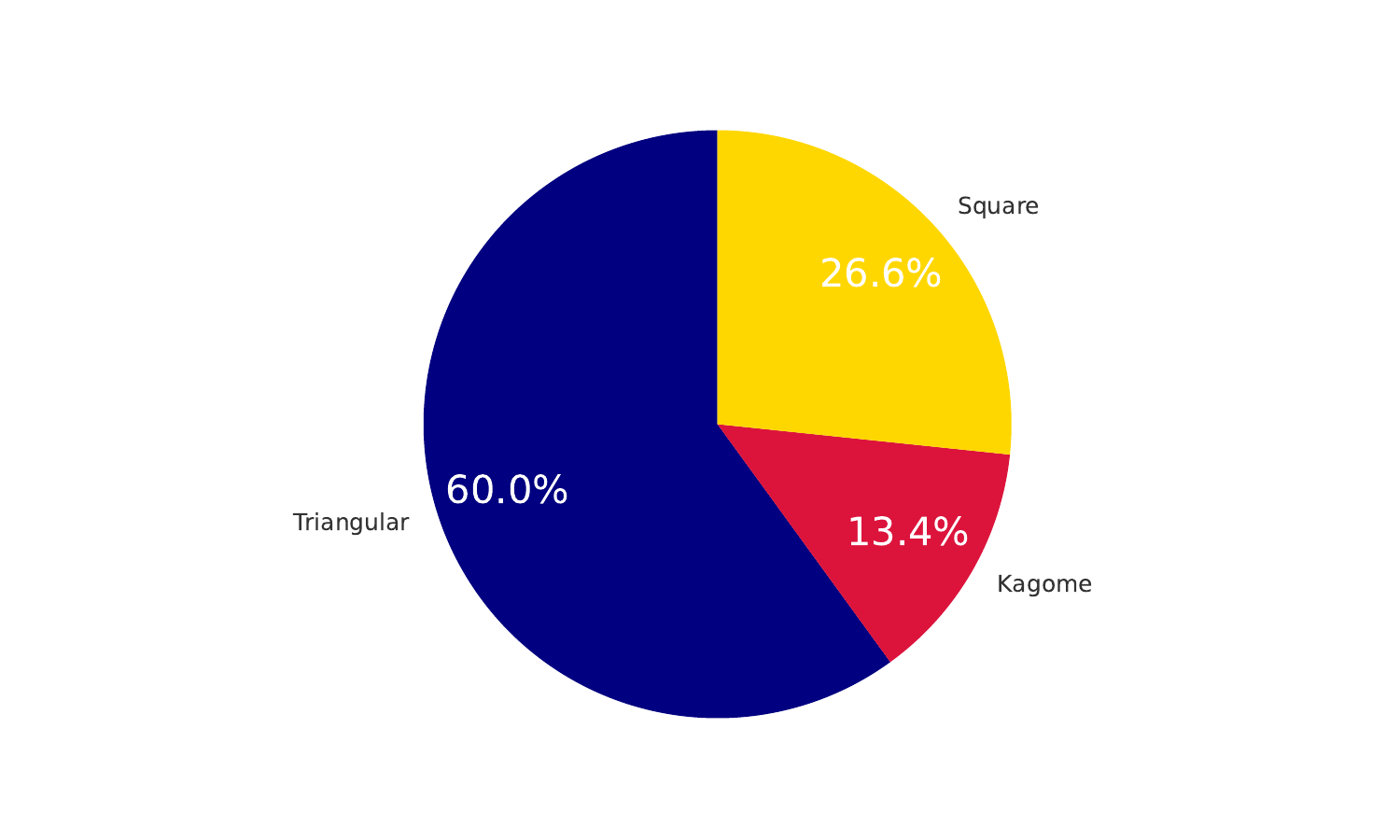}
\caption{Here we plot on which geometry the 10\% largest subgraphs created during the GLS mapping starting from for Erd\"os-Rényi graphs would be mapped. We observe that the largest subgraphs are more frequently mapped on Triangular lattices.}
\label{fig:sublattice_mapping}
\end{figure}

Finally, let us observe that the proposed  algorithm depends on the target lattice by construction.
We explored three different calibrated layouts on our quantum platform that could be used as target. As shown in Fig.~\ref{fig:sublattice_mapping}, we score the lattices in terms of where the 10\% largest subgraphs of Erd\"os-Rényi graphs would be mapped according to the GLS mapping.
We observe that in $60\%$ of cases, the largest subgraphs appear when the target lattice is triangular. For this reason, we chose the triangular lattice in our numerical simulations and experiments.

\section{Additional numerical results}\label{sec:additional_results}
\begin{figure}
    \includegraphics[width=\linewidth]{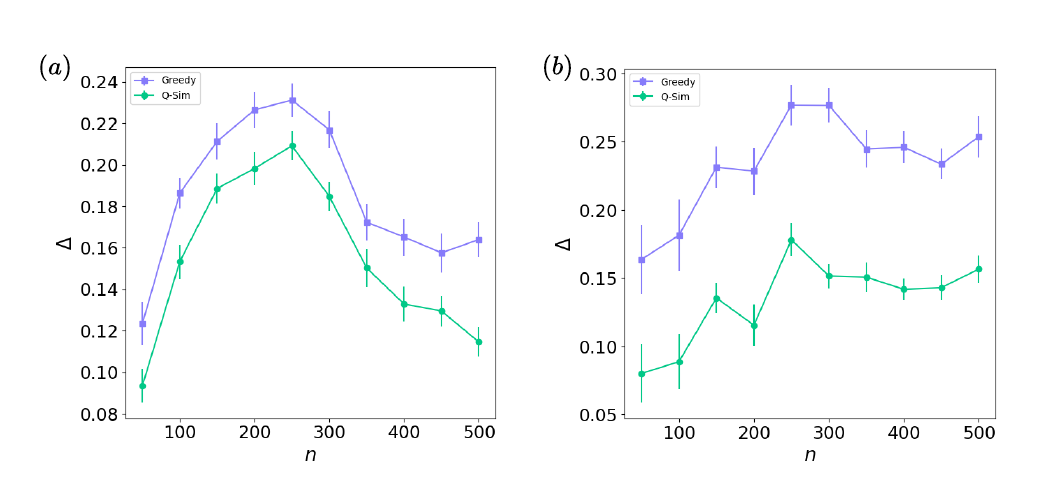}
    \caption{We plot the size of the MIS obtained via quantum simulations (Q-sim) by employing our method in comparison with the solution obtained by a Greedy algorithm, CPLEX~\cite{cplex2024} and Simulated annealing (SA), for Erd\"os-Rényi graphs of size up to $n=500$. We consider the probability of drawing an edge between two nodes as $(a)$ $p=0.25$; $(b)$ $p=0.75$. The results are averaged over 100 instances. The error bars correspond to the standard deviation of the mean.
    }
    \label{fig:supp_mat_results}
\end{figure}

In this section we present additional numerical results for the MIS problem on Erd\"os-Rényi graphs. We consider two different densities $p=0.25$ and $p=0.75$ that have not been presented in the main text.
The results are shown in Fig.~\ref{fig:supp_mat_results}.
We observe that the average value obtained by quantum simulations employing our method are always better than the Greedy counterpart. For completeness we plot both the error on the means and the standard deviation of the results. We observe that our method always outperforms the Greedy algorithm.

\end{document}